# Application of reciprocity for facilitation of wave field visualization and defect detection

Bernd Köhler[a*], Kanta Takahashi[b], Kazuyuki Nakahata[b]

*[a]Fraunhofer IKTS, Department Material Diagnostics, Dresden, Germany, [b]Ehime University, Civil and Environmental Engineering, Matsuyama, Japan*

* Fraunhofer IKTS, Maria-Reiche Str. 2, 01109, Dresden, Germany. Email: Bernd.koehler@ikts.fraunhofer.de

**Abstract:** The motion visualization in a structural component was studied for defect detection. Elastic motions were excited by hammer impacts at multiple points and received by an accelerometer at a fixed point. Reciprocity in elastodynamics is only valid under certain conditions. Its validity under given experimental conditions was derived from the elastodynamic reciprocity theorem. Based on this, the dynamic motion of the structural component was obtained for fixed-point excitation from measurements performed using multipoint excitations. In the visualized eigenmodes, significant additional deformation was observed at the wall thinning inserted as an artificial defect. To prevent the dependence of defect detection on its position within the mode shape, another approach was proposed based on the extraction of guided wave modes immediately after impact excitation. It is shown that this maximum intensity projection method works well in detecting defects.

**Keywords:** elastodynamic reciprocity; visualization of damage; multisite excitation; selective guided wave mode

## Introduction

Many attempts have been made to study the vibrations of civil engineering structures and use this information to schedule their maintenance. Some studies have identified the natural frequency of structures with high accuracy to characterize their state [1], whereas others have attempted to quantify the deterioration of structures from a decrease in their eigenfrequencies [2]. Furthermore, remote vibration monitoring [3, 4] is still being actively studied. A vibration measurement method using a portable sensor has been proposed for the quantitative evaluation of the soundness of small- and medium-sized bridges [5, 6, 7]. This method performs impact excitation at a certain point on the structure, measures the acceleration with sensors placed at multiple points on the surface of the structure, and calculates the displacement based on the data. However, it requires time and effort to install and replace multiple sensors. In practice, there are many problems in applying the above method. In addition to the need for many sensors, it is also necessary to synchronize the signals of all these sensors, requiring many cables and digitizing the signals in one device with a common time base for all channels. The requirement for simultaneous recording of signals from all points can be relaxed if the excitation is repeatable in its temporal characteristics and if a reliable reference time can be extracted for each excitation. A specific sensor can then be placed one after the other at various points for individual measurements. However, the sensor must be carefully placed and mechanically coupled for each measurement, which requires considerable time and effort.

An alternative to avoid sensor coupling issues is non-contact vibration monitoring. Known approaches include vision based optical techniques [8] scanning microphone detection [9] and Laser Doppler Vibrometry (LDV) [10-14]. All of these methods have their own challenges. The vision based technique and the microphone detection technique require fairly large vibration amplitudes. The LDV technique is widely applied for basic studies of elastic wave propagation in different materials such as concrete [11] and composites [12], and for studies of the dispersion relations of guided wave modes [13,14]. However, LDVs are very expensive and cannot easily be applied to real construction objects.

In contrast to measuring at multiple points, it is easy to detect elastic waves at a fixed point when the wave is generated at multiple points by an impact hammer. Impact hammer excitation is widely applied in various fields, for example fouling detection by vibration analysis [15] and modal analysis [16]. Therefore, we investigated the possibility of replacing the situation "scanning wavefield displacement detection with fixed-point hammer impact excitation" with the reverse situation "fixed-point displacement detection of wave fields excited by a scanning hammer impact".

The equivalence of the response when excitation and detection are reversed is normally denoted as reciprocity. A more precise term might be measurement reciprocity. While measurement reciprocity is often assumed, it is not always valid, as will be shown later. However, in the literature, its exact validity for a given measurement situation is often stated without proof. To illustrate, we provide give a few examples. Matsuoka [17] dealt with rails assuming measurement reciprocity in the sense that the excitation and measurement positions could be swapped, giving the same result. There is an integral equation in the cited literature [18] without an obvious connection to the measurement reciprocity of the system he considers. Lee et al. [19] focused on the degradation of sensors in structural health sensor networks. They state: "If one PZT transducer is used as an actuator and a second identical transducer is used as a sensor, and vice versa, the two measured time responses between them are reciprocal because of the reciprocity theorem for a linear system." This statement was not found in the cited literature [20], and it was again not clear how it could be obtained from the formulas given there. Yashiro et al. [21] demonstrated that scanning laser Doppler vibrometer measurements of wave propagation can be replaced by scanning laser excitation. They stated equivalence between the two situations: (a) excitation by a laser pulse on the surface in the thermoelastic regime at point "A" and detection by a piezoelectric probe at "B" and (b) excitation with the probe at "B" and detection at A. Through extensive numerical simulations, they demonstrated that such an equivalence is given. However, they used simplified assumptions regarding the source and neglected the influence of the transducer. A careful investigation of their simulation results shows small but significant deviations between some of the situations considered being reciprocal to each other (compare the early part in the time signal in Fig. 8 of [21]). Strict reciprocity between (a) thermoelastic laser excitation and piezoelectric detection and (b) piezoelectric excitation and out-of-plane displacement detection by a laser Doppler vibrometer (LDV) is not known and is generally not valid, according to our experience. To be clear at this point: we do not say that investigations under an unproven assumption of reciprocity must be wrong. In the published cases the reciprocity is valid or at least approximately valid and some of the results really open up new possibilities for NDE [21-26].

There are more general and clearly defined terms for "reciprocity." These are called reciprocity theorems and can be proved to be valid under defined conditions. Several such reciprocity theorems exist in elastodynamics [20] and electrodynamics [27]. Although not common in the literature on nondestructive testing, we argue that the measurement reciprocity for a given system should always be derived from such a reciprocity theorem.

The main objective of this study is therefore to derive the reciprocity between hammer impact excitation and displacement detection for a geometrically complex structure with given boundary conditions. We start from the reciprocity theorem [28] and incorporate boundary conditions describing the support of the structure in a simple model. We list different intermediate equations and arrive at statements about which quantities are reciprocal to each other. The intermediate equations given should help the reader to specify the derivation to more complex measurement situations of his experimental situation.

The other objective of this study is to apply the now-proven measurement reciprocity to facilitate the wavefield visualization and defect detection in a complicated structure by exchanging the positions of the hammer impact excitation and out-of-plane vibration detection. To the best of the authors´ knowledge, measurements based on that exchange have been performed only once so far for wavefield visualization for rails (see [17]) and never for defect detection. However, the boundary conditions of rails mounted on sleepers considered in [17] are complex and our derivation of measurement reciprocity does not apply to that case. As also the authors did not provide a prove, it is not clear whether reciprocity holds strictly or at least approximately in their case.

To keep the focus of the manuscript, the proposed method for defect identification is demonstrated only for detection and location of a simple model defect. Its evaluation for other defect types and defect sizing is not within the scope of the paper. The application of measurement reciprocity for other methods of defect characterization is discussed shortly at the end of the paper.

The remainder of this paper is organized as follows. First, the measurement reciprocity for point force excitation and point displacement detection is derived. A special type of component support is also included, and it is shown that it does not deviate from reciprocity. Subsequently, we describe the investigated component (the I-beam), our measurement setup, and the applied signal-processing scheme. We also experimentally demonstrate measurement reciprocity and show that reciprocity is not valid for arbitrary displacement components – the displacement components have to be selected according to the measurement reciprocity derived from the reciprocity theorem. With this reciprocity, we replace the multipoint measurement with multipoint excitation to facilitate the measurement, particularly in field applications. We first apply the reciprocal measurement to wavefield visualization. Next, we use these tools to get vibration data in a test I-beam with an artificial defect in the form of thinning. The obtained data are used to visualize the 3D motion of the beam. The transient motion shortly after the impact excitation and eigenmodes are shown. We demonstrate that a specific eigenmode can mark the area with an artificial defect. Next, we propose a further method based on guided wave modes early after excitation to overcome some disadvantages of the eigenmode method. Finally, we summarize the study and suggest possible further work.

# Reciprocity for point-excited vibration measurements

## *General considerations*

We begin with the time-dependent reciprocity relation as formulated in [28]. We use this relation in the following to the experimental situation described later in this study. We consider an elastic body with volume V, surface S, and time-invariant properties. In general, there can be volume forces $\boldsymbol{f}(\boldsymbol{r}, t)$ inside V and surface tractions $\boldsymbol{t}(\tilde{\boldsymbol{r}}, t) = \boldsymbol{\sigma} \cdot \boldsymbol{n}$ acting on S. Here, $\boldsymbol{r}$ denotes the position vector in volume V, $\tilde{\boldsymbol{r}}$ is the position vector $\boldsymbol{r}$ in surface S, $\boldsymbol{\sigma}$ is the stress tensor, and $\boldsymbol{n}$ *is* an outward directed surface normal vector. The corresponding displacement field is $\boldsymbol{u}(\boldsymbol{r}, t)$. Next, there holds a reciprocity theorem (Equation (5.15) in [28]) for two admissible elastodynamic states denoted as $\boldsymbol{u}^a, \boldsymbol{f}^a, \boldsymbol{t}^a$ and $\boldsymbol{u}^b, \boldsymbol{f}^b, \boldsymbol{t}^b$. In our case, there are only surface tractions on S but no body forces $\boldsymbol{f}$. Furthermore, we assume that the body is at rest for all times $t \leq 0$ and that surface tractions act only at time $t > 0$. Then, for displacements $\boldsymbol{u}(\boldsymbol{r}, t)$ and velocities $\dot{\boldsymbol{u}}(\boldsymbol{r}, t)$, we can set

$$\boldsymbol{u}^a(\boldsymbol{r}, 0^+) = 0, \qquad \dot{\boldsymbol{u}}^a(\boldsymbol{r}, 0^+) = 0, \qquad \boldsymbol{u}^b(\boldsymbol{r}, 0^+) = 0, \; \dot{\boldsymbol{u}}^b(\boldsymbol{r}, 0^+) = 0$$

Under these assumptions, the reciprocity theorem (5.15) of [28] takes the form

$$\int_S \boldsymbol{t}^a * \boldsymbol{u}^b \, dS = \int_S \boldsymbol{t}^b * \boldsymbol{u}^a \, dS$$

or in the Cartesian components

$$\int_S t_j^a(\boldsymbol{r}, t) * u_j^b(\boldsymbol{r}, t) \, dS = \int_S t_j^b(\boldsymbol{r}, t) * u_j^a(\boldsymbol{r}, t) \, dS \qquad (1)$$

Note that we used the symbols (a, b, $\boldsymbol{r}$) for the two states and the position vector, whereas in [28], the symbols (A, B, $\boldsymbol{x}$) were used. A repeated index implies a summation over the range of that index (typically 1–3). The asterisk in (1) denotes a convolution in the time domain of the form

$$f_1(t) * f_2(t) = \int_0^t f_1(t - s) f_2(s) \, ds. \qquad (2)$$

For later use, we note that the convolution is commutative, that is, $(f_1 * f_2)(t) = (f_2 * f_1)(t)$.

Consider a typical experimental scenario (Figure 1). A specimen is supported somewhere (here, at points SP1 and SP2), exited at point A with coordinates $\boldsymbol{r}_A$, and the response is measured at point B. We denote the corresponding state as state "a." We can ensure that the measurement of the surface displacement does not introduce surface tractions in many experimental situations. This is particularly true for LDV measurements but will also be a good approximation for lightweight acceleration sensors. The surface tractions $\boldsymbol{t}^a$ are zero everywhere, except at the support points (SP) and excitation point A. The integral in the left-hand side of (1) thus splits up into several non-zero contributions, with the additional summand $(+\ldots)$ indicating contributions of possibly existing further support points:

$$I^{lhs} = I_A^{lhs} + I_{SP1}^{lhs} + I_{SP2}^{lhs} + \cdots$$

$$= \int_{S1=\delta A} \ldots + \int_{S2=\delta SP1} \ldots + \int_{S3=\delta SP2} \ldots + \cdots \tag{3}$$

There is a corresponding expression for the right-hand side of (1)

$$I^{rhs} = I_B^{rhs} + I_{SP1}^{rhs} + I_{SP2}^{rhs} + \cdots$$

$$= \int_{S1=\delta B} \ldots + \int_{S2=\delta SP1} \ldots + \int_{S3=\delta SP2} \ldots + \cdots \tag{4}$$

The "$\delta$," e.g., in $\delta A$, indicates that the integration should be done over a small area around point A on surface S.

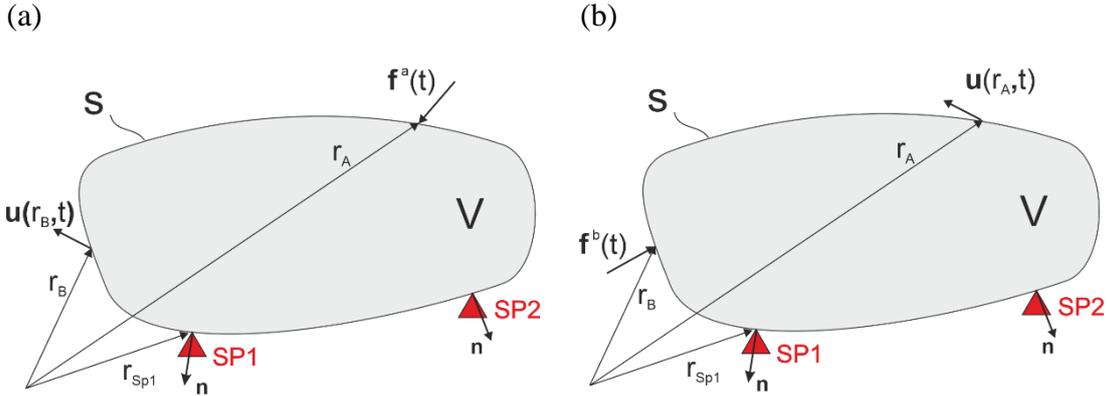

**Figure 1.** Volume V with surface S. Volume is assumed free from body forces, and the surface is assumed traction-less except at support points SP and at one additional point where the wave motion is excited by a point force acting at t > 0. At SP, we assume a point contact and that only normal forces are transferred (that is, sliding is allowed at support points).
left: state (a)—excitation at point A by a point force $\boldsymbol{f}^a(t)$;
right: state (b)—excitation at point B by a point force $\boldsymbol{f}^b(t)$.

In many experimental situations, the excitation at the surface can be approximated by a point force acting in a certain direction $\boldsymbol{e}$ with a time dependence $f(t)$: $\boldsymbol{f}(t) = \boldsymbol{e}f(t)$, i.e. the force direction $\boldsymbol{e}$ does not depend on time. We restrict the following to such point force excitations. Then the surface tractions for states "a" and "b" are

$$\boldsymbol{t}^a(\tilde{\boldsymbol{r}},t) = \boldsymbol{e}^a \, f^a(t)\delta^2(\tilde{\boldsymbol{r}} - \tilde{\boldsymbol{r}}_A) \quad \text{and} \quad \boldsymbol{t}^b(\tilde{\boldsymbol{r}},t) = \boldsymbol{e}^b \, f^b(t)\delta^2(\tilde{\boldsymbol{r}} - \tilde{\boldsymbol{r}}_B)$$

where, again, the "tilde" on symbol $\tilde{\boldsymbol{r}}$ indicates the restriction of $\boldsymbol{r}$ to surface S (see [29] for the concept of the surface traction). When it is clear from the context that $\boldsymbol{r}$ is in S, the tilde is also omitted. Therefore, we have $\tilde{\boldsymbol{r}}_A = \boldsymbol{r}_A$. The $\delta^2()$ denotes the two-dimensional Dirac delta function on S.

The first integrals in (3) and (4) simplify to

$$I_A^{lhs} = \int_{S1=\delta A} t_j^a(\boldsymbol{r},t) * u_j^b(\boldsymbol{r},t)dS = f^a(t) * e_j^a u_j^b(\boldsymbol{r_A},t) \tag{5}$$

and

$$I_B^{rhs} = \int_{S1=\delta B} t_j^b(\boldsymbol{r},t) * u_j^a(\boldsymbol{r},t)dS = f^b(t) * e_j^b u_j^a(\boldsymbol{r_B},t) \tag{6}$$

Next, we consider the contributions of SPs for which we cannot assume vanishing traction. However, no in-plane traction is typically transferred when the specimen rests on the SPs. Therefore, we can assume that $\boldsymbol{t(n)}$ is parallel to $\boldsymbol{n}$ for all the SPs. Moreover, it is reasonable to assume that traction is point like and proportional to the displacement normal to the surface (that is, we have spring contact with a spring constant k). Next, we consider SP1. For the tractions in states "a" and "b," we have at SP1:

$$\boldsymbol{t}^a(\tilde{\boldsymbol{r}},t) = -\boldsymbol{n} \cdot \boldsymbol{u}^a(\tilde{\boldsymbol{r}},t)k\delta^2(\tilde{\boldsymbol{r}} - \tilde{\boldsymbol{r}}_{SP1})\boldsymbol{n} \ \text{ and } \ \boldsymbol{t}^b(\tilde{\boldsymbol{r}},t) = -\boldsymbol{n} \cdot \boldsymbol{u}^b(\tilde{\boldsymbol{r}},t)k\delta^2(\tilde{\boldsymbol{r}} - \tilde{\boldsymbol{r}}_{SP1})\boldsymbol{n}$$

Therefore, the contributions to the integrals from SP1 equate to

$$I_{SP1}^{lhs} = \int_{\delta SP1} \{-\boldsymbol{n} \cdot \boldsymbol{u}^a(\tilde{\boldsymbol{r}},t)\}k\delta^2(\tilde{\boldsymbol{r}} - \tilde{\boldsymbol{r}}_{SP1})\boldsymbol{n} * \boldsymbol{u}^b(\tilde{\boldsymbol{r}},t)dS$$

$$= -k(\boldsymbol{n} \cdot \boldsymbol{u}^a(\tilde{\boldsymbol{r}}_{SP1},t)) * (\boldsymbol{n} \cdot \boldsymbol{u}^b(\tilde{\boldsymbol{r}}_{SP1},t))$$

and

$$I_{SP1}^{rhs} = \int_{\delta SP1} \{-\boldsymbol{n} \cdot \boldsymbol{u}^b(\tilde{\boldsymbol{r}},t)\}k\delta^2(\tilde{\boldsymbol{r}} - \tilde{\boldsymbol{r}}_{SP1})\boldsymbol{n} * \boldsymbol{u}^a(\tilde{\boldsymbol{r}},t)dS$$

$$= -k(\boldsymbol{n} \cdot \boldsymbol{u}^b(\tilde{\boldsymbol{r}}_{SP1},t)) * (\boldsymbol{n} \cdot \boldsymbol{u}^a(\tilde{\boldsymbol{r}}_{SP1},t))$$

The convolution * is commutative; therefore, both integrals are equal ($I_{SP1}^{lhs} = I_{SP1}^{rhs}$). As we can repeat the argument for the other SPs, all the contributions of SPs on the left-hand side in (1) are equal to their contributions on the right-hand side. Therefore, in (1), only the terms $I_A^{lhs}$ and $I_B^{rhs}$ remain, giving $I_A^{lhs} = I_B^{rhs}$ and with (5) and (6)

$$f^a(t) * e_j^a u_j^b(\boldsymbol{r_A},t) = f^b(t) * e_j^b u_j^a(\boldsymbol{r_B},t). \tag{7}$$

Equation (7) describes the reciprocity theorem in the time domain when specified to the experimental situation of Figure 1. To be clear we repeat the assumptions and statements of the theorem. Assume there is an elastic body which is traction free except at a few support points. The support point tractions can be described using spring forces normal to the surface. Two elastodynamic states "a" and "b" are considered. The displacement field $\boldsymbol{u}^a(\boldsymbol{r}, t)$ of the state "a" is excited by a point force acting on the surface at $\boldsymbol{r} = \boldsymbol{r_A}$ (point A) in direction $\boldsymbol{e}^a$ with the time dependence $f^a(t)$. Correspondingly, the displacement field $\boldsymbol{u}^b(\boldsymbol{r}, t)$ of the state "b" is excited by a point force acting at $\boldsymbol{r} = \boldsymbol{r_B}$ (point B) in direction $\boldsymbol{e}^b$ with the time dependence $f^b(t)$. The theorem relates the surface displacements of both states at the excitation points of the corresponding reciprocal states. That is for the state "a" the displacement is taken at the point B and for the state "b" it is taken at point A. The theorem (equation (7)) can be phrased as follows: the state "a" displacement $\boldsymbol{u}^a$ projected to $\boldsymbol{e}^b$ (the reciprocal state excitation force unit vector) and additional convolved with the time history of the reciprocal state $f^b(t)$ is equal to the same value calculated for both states exchanged.

We specify this reciprocity theorem further in the following section for impact excitation.

### Special cases: impact excitation and identical time course of the excitation forces

The impact is a very short event and can often be approximated as a Dirac delta function. To specify (7) for this case, we assume that

$$f^a(t) = f_0^a \delta(t - t_a), \; f^b(t) = f_0^b \delta(t - t_b)$$

where both impact times are positive: $t_a > 0$ and $t_b > 0$. Note that $t_a$ and $t_b$ of the impacts in states "a" and "b" must not necessarily be the same. Equation (7) then becomes

$$e_j^a \hat{u}_j^b(\boldsymbol{r_A}, t - t_a)/f_0^b = e_j^b \hat{u}_j^a(\boldsymbol{r_B}, t - t_b)/f_0^a \qquad (8)$$

where, the "^" symbol on $\hat{u}$ indicates that the displacement is the response to a $\delta$-excitation. In the special case where the $\delta$-excitations in both states have equal amplitudes and times ($f_0^a = f_0^b, t_a = t_b$), however, still not necessarily the same orientation, we obtain a further simplified version:

$$e_j^a \hat{u}_j^b(\boldsymbol{r_A}, t) = e_j^b \hat{u}_j^a(\boldsymbol{r_B}, t). \qquad (9)$$

The excitation is not always very short and unipolar to be approximated by a Dirac-$\delta$. We assume that the time histories of both the excitation forces are equal. From the linear superposition, the solution $\boldsymbol{u}(t)$ for excitation with force $\boldsymbol{f}(t) = f(t) f_0 \boldsymbol{e}$ can be described as the convolution of the solution $\hat{\boldsymbol{u}}(t)$ for excitation $\boldsymbol{f}(t) = f_0 \delta(t - t_0) \boldsymbol{e}$ with the excitation time course $f(t)$, that is, $\boldsymbol{u}^{a/b}(t) = f(t) * \hat{\boldsymbol{u}}^{a/b}(t)$. By convolving both sides of (9) with $f(t)$, we obtain

$$e_j^a u_j^b(\boldsymbol{r_A}, t) = e_j^b u_j^a(\boldsymbol{r_B}, t) \qquad (10)$$

which is the same as (9) except that fields $\boldsymbol{u}^{a/b}$ are now the solutions for the excitation of a common but arbitrary time course $f(t)$. It is worth mentioning that in this equation from field $\boldsymbol{u}^b$, only the projection in the direction of force $\boldsymbol{f}^a$ is taken and vice versa.

We consider the excitation using normal forces as the last special case. We are most interested in this case because the hammer mainly transmits normal forces. Specifying in (10) the unit vector $\boldsymbol{e}$ of the force to the surface normal unit vector $\boldsymbol{n}$, we obtain

$$\boldsymbol{n}^a \cdot \boldsymbol{u}^b(\boldsymbol{r}_A, t) = \boldsymbol{n}^b \cdot \boldsymbol{u}^a(\boldsymbol{r}_B, t). \tag{11}$$

Let us summarize the assumptions leading to (11):
a)  the specimen has time-independent linear elastic properties,
b)  there are no body forces,
c)  the sensors do not introduce significant tractions to the surface (no reactance),
d)  point-like SPs introduce only point forces normal to the surface proportional to the displacement and
e)  the excitation is a point force normal to the surface, and its time course is the same for both states (i.e., at both positions A and B).

Assumption (e) is not trivial and must be verified. However, for our experimentational conditions (e) is fulfilled as described in the next section.

## Experimental methods

### *Hardware*

In guided-wave excitation, an ultrasonic probe with a limited frequency band is often used. Here, we consider the hammer excitation with a rather high relative bandwidth. This has the additional advantage that we can change the excitation point quickly without the need for a careful setup of good coupling. We constructed a measurement system using wireless communication, which reduces the number of cables to apply the setup to large structures. This system contains wireless sensor nodes consisting of a sensor, an A/D converter, a base station with a wireless LAN router, and a notebook PC. In this study we connected both, the impact hammer and the three-dimensional acceleration sensor to a single wireless device. In applications where the excitation and detection positions are more distant from each other, it is possible reducing the cabling by using two wireless devices, as demonstrated in [7].

The impact hammer FHA2KC, manufactured by Fuji Ceramics Co. Ltd., has a mass of 180 g. Excitation force can be measured with an internal sensor with a sensitivity of 2.2 mV/N. The maximum allowable load is 2225 N. A resonance frequency of 31 kHz limits the usable frequency range to 8 kHz. The frequency content of the impact excitation can be adjusted by selecting the hammer tip. A stainless steel tip was selected to excite a broad frequency band, including high-frequency components. The excitation points were marked by a ruler. The precision of the excitation position is estimated to be $\pm$ 1 mm.

Figure 2 shows an example of the force versus time trace generated by a hit on the sample, together with the corresponding Fourier spectrum. Hits at various points on the structure always showed the same force–time trace and spectrum, and only the amplitude varied from hit to hit owing to manual operation. The received signals were scaled to the maximum excitation force to remove any variation. The observation that the force–time trace does not depend on the excitation point can be expected: the active mass of the hammer (180 g) is small compared to the mass of the sample around all excitation points. Therefore, the necessary condition (e) for measurement reciprocity ((10) and (11)) is also fulfilled.

A lightweight piezoelectric 3-axis acceleration sensor (SA11ZSCA, Fuji Ceramics Co.) was used as receiver. Its mass is only 4.4 g. The sensor is specified for accelerations up to 5 000 m/s² with a sensitivity of 2.2 mV/(m/s²). The lower limit of the specified frequency range is 0.5 Hz for all three axes. The upper limit differs and is 15 and 20 kHz for the in-plane and out-of-plane orientations, respectively. The acceleration sensor was attached to the specimen using silicon grease.

The voltage signals from the sensor and impact hammer were digitized using an A/D converter from National Instruments (NI9215). It had four channels with a resolution of 16 bits, voltage input range of ±10 V, and maximum sampling rate per channel of 100 kS/s. The sensible sampling rate setting depends on the highest frequencies excited. The spectrum was determined by the hammer's internal sensor; based on this, the sampling rate was set to 51.2 kHz. The converted digital signal was sent to a notebook PC via the wireless LAN router WLS/ENET-9163 (National Instruments), providing a maximum transfer rate of 5MS/s Time synchronization of the sequentially recorded signals at all measurement points is important for 3D visualization. In this study, the acceleration pulse signal generated by the internal sensor of the impact hammer was used as a "soft trigger" to synchronize the recordings from different positions.

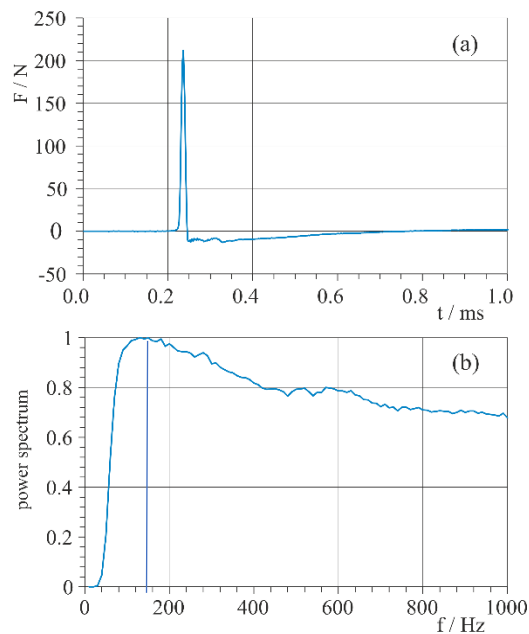

**Figure 2.** (a) Force F(t) exerted by the impact hammer on the surface measured by its internal sensor; (b) the corresponding normalized power spectrum.

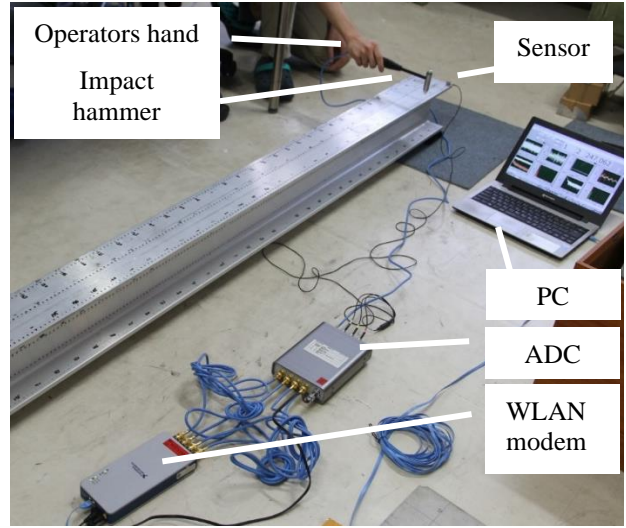

**Figure 3.** Image showing some of the used components in operation on an I-beam.

### *Calculation of displacement*

The following linear acceleration method [30] was used to convert the acceleration to displacement. Let $\Delta t$ be the sampling interval and n be an integer representing the current time step for $\boldsymbol{v}^n = (v_x^n, v_y^n, v_z^n)$ and $\boldsymbol{a}^n = (a_x^n, a_y^n, a_z^n)$.

The velocity at step n is calculated as

$$\boldsymbol{v}^n = \boldsymbol{v}^{n-1} + \Delta\text{t}\left[\frac{\boldsymbol{a}^{n-1}}{2} + \frac{\boldsymbol{a}^n}{2}\right] \tag{12}$$

and the displacement as

$$\boldsymbol{u}^n = \boldsymbol{u}^{n-1} + \Delta\text{t}\boldsymbol{v}^{n-1} + \Delta\text{t}^2\left[\frac{\boldsymbol{a}^{n-1}}{3} + \frac{\boldsymbol{a}^n}{6}\right] \tag{13}$$

where $\boldsymbol{v}^0 = \boldsymbol{0}$ and $\boldsymbol{u}^0 = \boldsymbol{0}$. As Equations (12) and (13) describe a numerical integration, there may be deviations in the baseline of the calculated displacement. This is known as the displacement drift error [31] and can be corrected using a digital filter. In this study, a 10-Hz-high-pass digital filter was used to remove only low frequencies in the signal that cause drift errors.

### *Experimental demonstration of measurement reciprocity*

The reciprocity theorem is exact and does not require verification. However, its validity depends on the fulfilment of the prerequisites. Therefore, it is a good idea to demonstrate reciprocity for a given setup by comparing the time responses when the excitation and reception points are exchanged. This was performed on an I-beam with a cross-section similar to that used later. The specimen was made of aluminum with a longitudinal wave speed of 6360 m/s, shear wave speed of 3100 m/s, and density of 2700 kg/m³. The beam 2 m in length was supported by four point-like contacts at its ends.

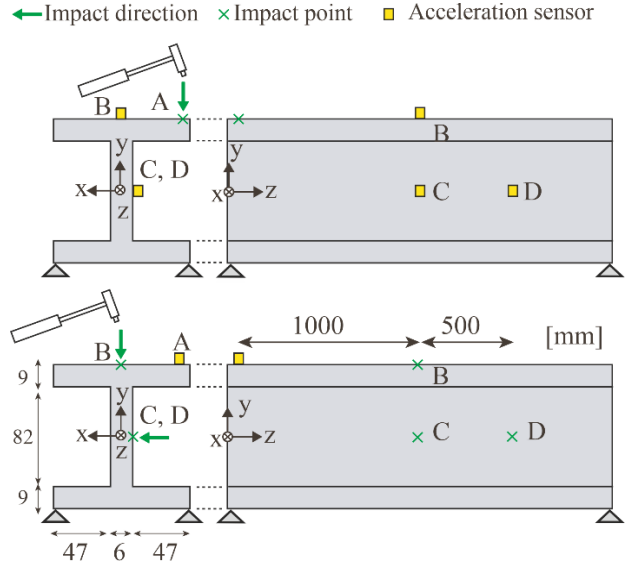

**Figure 4.** Scheme of sample for reciprocity evaluation (not at scale). Green crosses depict the excitation-impact points. Sensor position is given as a yellow square. All distances are in mm. Top image: state a–excitation is at A, and the sensing is at one of the points B, C, and D; Bottom image: state b—sensing is at A, and the excitation is at one of the points B, C, and D.

Reciprocity between pairs of exciter–detector locations was tested. One of the locations was always named A. Depending on the experiment, the other location of the pair was B, C, or D. The locations were as follows:

A is the endpoint of the upper flange at z = 0 mm,
B is the center of the upper flange at z = 1000 mm,
C is the center of the web at z = 1000 mm, and
D is the center of the web at z = 1500 mm.

In state (a) (top of Figure 4), vibration is impact excited at A, and the signal is recorded at one of the locations B, C, or D, whereas in state (b) (bottom of Figure 4), the signal is measured at location A and excited at one of the other points. The results of these measurements are shown in Figure 5. The blue lines always show the waveform in state (a), whereas the dashed red lines show the waveform when the excitation and reception points are reversed, that is, in state (b).

As mentioned, the sensor used was a 3D accelerometer; therefore, acceleration $\boldsymbol{a}$ was measured in all three spatial directions. From the numerical integration (13), we also obtain the full vector of displacement $\boldsymbol{u}$. Equation (11) indicates which components of the measured displacements are reciprocal to each other for states (a) and (b). These are always projections of the displacement vector onto the excitation force vector of the complementary state. For other components, there is simply no statement that can be derived from the theorem. In our case, we only had excitations normal to the surface. Therefore, we used the displacement component, which was also normal to the surface at the corresponding points, that is, the components ($\boldsymbol{u}_y, \boldsymbol{u}_y, \boldsymbol{u}_x$, and $\boldsymbol{u}_x$) at points (A, B, C, and D), respectively.

The agreement between the displacement components selected according to this rule was very good ( Figure 5 left column). To illustrate the importance of the correct choice of the displacement component, we present counterexamples in the right column in Figure 5. For displacement components, for which we could not derive measurement reciprocity, there are indeed large discrepancies between the signals.

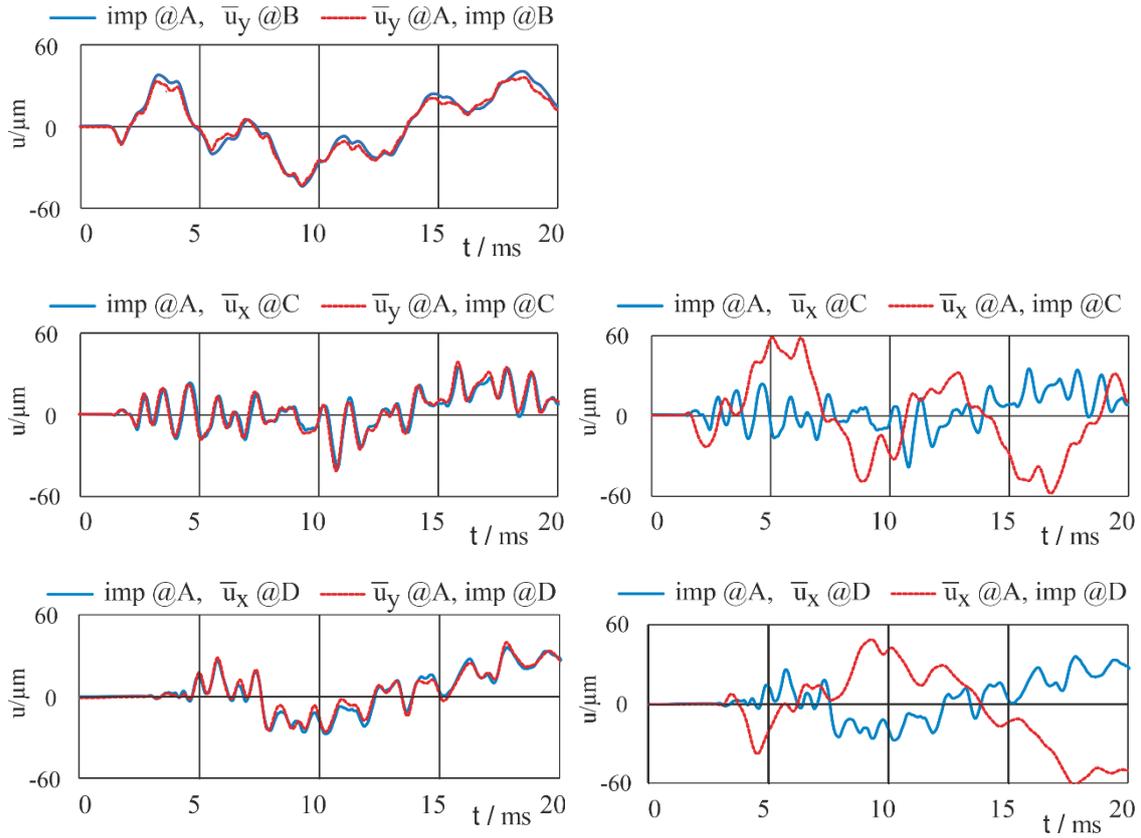

**Figure 5.** Displacement waveforms in the states obtained by exchange of impact excitation and detection locations. The pairs of locations are A-B, A-C, and A-D from top to bottom. The blue curve always corresponds to impact excitation at point A, whereas the red dotted curve corresponds to detection at A. In the right column, displacement components were selected for which we cannot derive measurement reciprocity. Not surprisingly, there are large discrepancies between the signals.

## 3D Visualization of dynamic behavior of structures

### *Specimen and measurements*

One objective of this study was to visualize the structure deformation after excitation at a given point using measurement reciprocity. The sample was identical to that shown in Figure 4 except for a thickness reduction over a length of 50 mm with a depth of 4.2 mm in the lower flange (see Figure 6). An acceleration sensor was mounted on the upper flange at a fixed position as the receiver. The excitation by hammer was performed along measurement lines (MLs) with a constant spacing of 10 mm between excitation points. Two hundred excitation points were used per line. Eight MLs were used in the study.

Owing to the proven measurement reciprocity, the measured data are equal to the (out of plane component of the) wavefield excited by hammer impact. In the reciprocal situation the excitation is at the position where in the actual measurement the acceleration sensor is mounted (the yellow square in Figure 6).

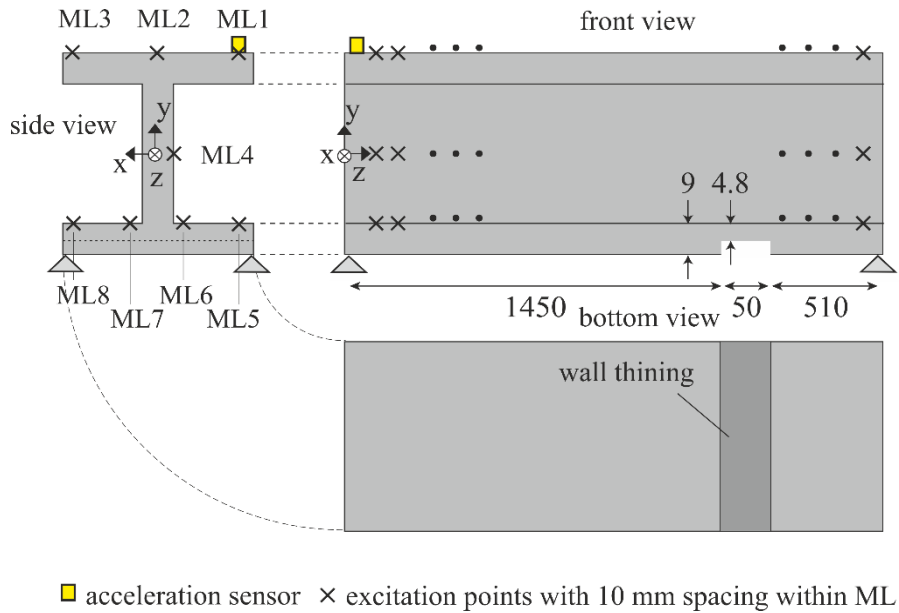

ML3  ML2  ML1          front view

side view

ML4

9  4.8

ML8                    1450  bottom view  50  510
ML7  ML6  ML5

wall thining

☐ acceleration sensor  × excitation points with 10 mm spacing within ML

**Figure 6**. Scheme of the I-shaped cross-section specimen (not at scale). All dimensions are in mm. The position of the fixed acceleration sensor and hammer excitation points with a spacing of 10 mm along the 8 measuring lines (ML1 … ML8) are represented by a yellow square and black crosses, respectively.

The obtained displacements are visualized in Figure 7 for a series of times after excitation. The color map represents the absolute displacement value |u| normalized to its maximum value. The wave propagation is evident in the figure. Multiple modes propagate in an overlapping state immediately after excitation; however, owing to different group velocities (energy propagation velocities), the modes become separated as time elapses. As the wave propagates and reaches the edge of the specimen, a reflection appears. Finally, after multiple reflections, a steady state of overlapping eigenmodes results.

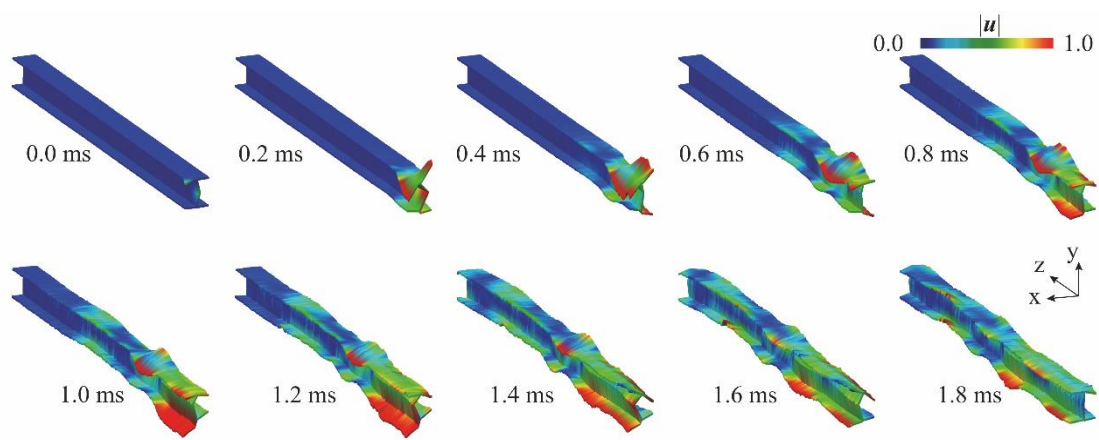

**Figure 7.** Snapshots of I-beam specimen deformation at various times shortly after excitation. This visualization is created by AVS/Express based on measured data.

*Extraction of natural vibration modes*

Because energy is trapped in the sample and the damping is rather low, a steady state forms after multiple reflections of the wave. Characteristics of the eigenmodes can be used as an indication of defects. To begin with the frequency characteristics, Figure 8 shows the spectra of the y-displacement along line ML1 on the upper flange (Figure 8a) and along line ML5 on the lower flange (Figure 8b). This type of presentation corresponds to frequency-domain B-scans [32] in ultrasonic testing. The pair of horizontal white dashed lines in Figure 8b indicates the artificial thinning position on the lower flange. The spectra of the scans of the upper (Figure 8a) and lower flanges (Figure 8b) are very similar. This is no surprise because the specimen is nearly symmetric and the scan lines are equivalent. The only difference is the thinning of the lower flange (Figure 6). At the location of thinning, the lower flange scan shows increased spectral values for frequencies corresponding to the first horizontal and second vertical bending modes. In the inspection method proposed in the following section, we focus on the second vertical bending mode because we expect this mode to have a better spatial resolution because of its higher frequency compared to that of the first horizontal bending mode.

The displacement filtered for the second vertical bending mode (428–442 Hz, 4th-order Butterworth) is shown in Figure 9. The second vertical bending mode is evident, and the thinned part shows a significantly increased displacement, which is expected owing to the increased spectral values at that position (see Figure 8b) [33]. This increase was not visible in the defect-free upper part of the beam, as shown in Figure 9 and Figure 8a.

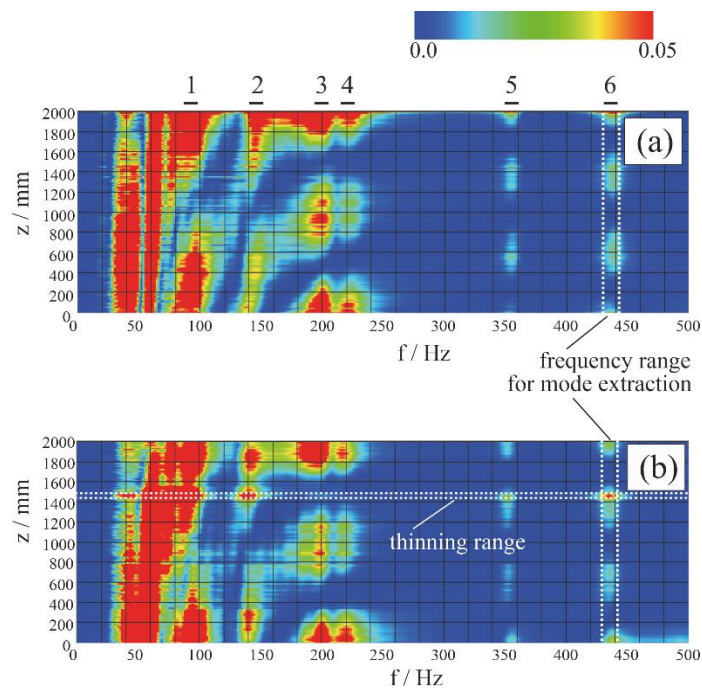

**Figure 8.** Fourier spectra of measured signals along (a) ML1 and (b) ML5.
The numbered bars on top of image (a) correspond to the following eigenmodes:
1: 1st torsion; 2: 1st horizontal bending; 3: vertical bending; 4: 2nd torsion 5: 3rd torsion; 6: 2nd vertical bending.

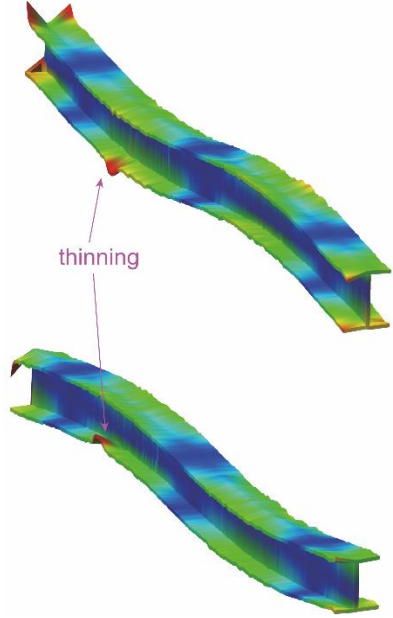

**Figure 9.** Visualization of displacement data after bandpass filtering around the natural frequency of the 2ⁿᵈ bending eigenmode ($f_{low} = 428$ Hz, $f_{high} = 442$ Hz). The eigenmode is clearly recognized.

## Damage detection from guided wave modes

The use of eigenmodes for defect characterization has two disadvantages. First, there is no defect indication when the defect position coincides with the node of the considered eigenmode. Second and more importantly, in practice we are dealing with large and complicated structures rather than a single beam of finite length. Thus, there will be no such pronounced eigenmodes because energy is not restricted to the location of excitation but propagates through the rest of the structure.

To address these challenges, we propose an approach that is new to the best of our knowledge. The proposed approach is based on similarity between guided wave modes and eigenmodes. In fact, eigenmodes in a structure of length $L$ can be considered guided wave modes with wavelength $\lambda$ determined as $L = (2n - 1)\lambda/2$. If the excitation point is not significantly far away from the area we are interested in (the surroundings of the defect), then it should be possible to isolate the guided wave mode of interest for times early after excitation. To follow this idea, we first study the guided wave modes both experimentally and numerically and then describe the extraction of information from the guided wave in an early stage after excitation.

### *Dispersion curve*

Analytical solutions exist for dispersion diagrams of simple waveguides such as plates and roads [34]. In our more complicated structure, we must calculate dispersion diagrams using numerical methods. A rather efficient method is called SAFE [31]. SAFE expresses ultrasonic waves propagating in the longitudinal direction (waveguide propagation direction) using a complex amplitude in the frequency domain f (phasor representation), divides the cross-sectional shape into elements, and reduces it to an eigenvalue problem. Finally, eigenwave numbers $\zeta_i$ (i = 1, 2, …) of wave propagation are obtained; thus, phase velocity $c_p$ for mode $i$ can be calculated from $c_{p,i} = 2\pi f / \zeta_i$. We implemented SAFE with quadrangular 8-node isoparametric elements. To obtain accurate results, the nodal spacing must be a factor of 5 smaller than the smallest wavelength for the frequencies of interest. With a nodal spacing of 2 mm 468 nodes in the cross section resulted.

The measured displacement values $u(z,t)$ were 2D-Fourier transformed into the wavenumber- frequency domain

$$H(k,f) = \int_{-\infty}^{\infty} \int_{-\infty}^{\infty} u(z,t) e^{i(kz - 2\pi ft)} dz \ dt. \qquad (14)$$

This transformation is performed as a 2D fast Fourier transformation. Therefore, $H$ is given for discrete values of frequency $f$ and wave number $k$. For good visualization, $H$ was normalized to its maximum for each fixed frequency $f_j$:

$$\overline{H}(k,f_j) = \frac{H(k,f_j)}{\max\limits_{0 \le k \le 0.1} H(k,f_j)} \qquad (15)$$

Figure 10 shows color-coded spectral values $\overline{H}(k,f)$ for ML1, ML4, and ML5. The dispersion curves obtained by SAFE are inserted in all three plots as white dots. Figure 10(a) and Figure 10(c) were calculated from the y-direction displacements of ML1 and ML5, and Figure 10(b) from the x-direction displacement of ML4.

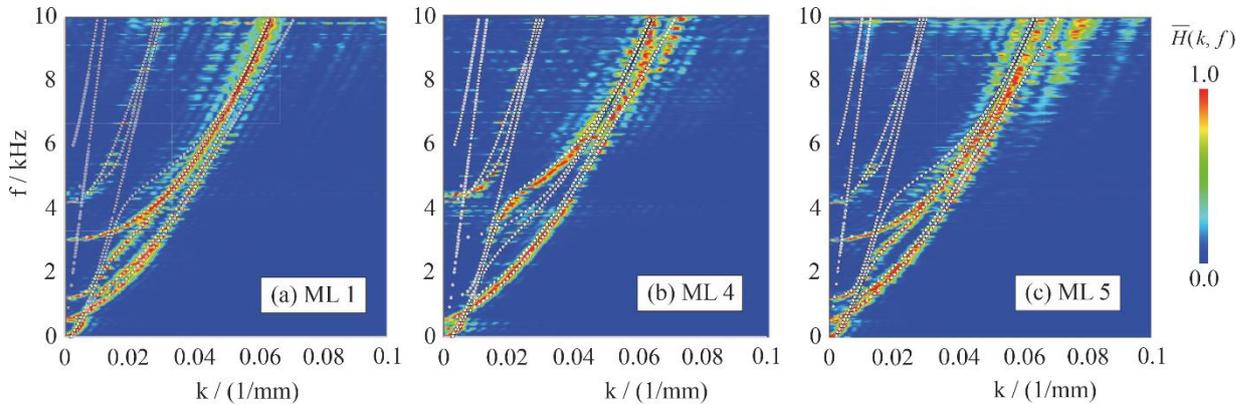

**Figure 10.** Color-coded 2D spectral data obtained from (out-of-plane) displacement measurements along (a) ML1, (b) ML4, and (c) ML5. In each 2D spectrum, the dispersion curves calculated using SAFE are displayed as white dots.

The high spectral values of the measured data coincided with the points on the numerically obtained dispersion curves. This agreement was rather good. The population of dispersion curves with high spectral values differed between the three graphs. This can be easily explained by the fact that each graph represents the out-of-plane displacement of a mode along a certain line. Some modes simply have no significant displacement at some lines, whereas strong displacements occur along others. A good example is the torsional mode around the center of the beam cross section, which leaves the middle of the web (ML4) with only negligible displacement. Some dispersion curves determined with SAFE show no agreement at all with significant spectral values from experiment. These modes were probably not excited by the impulse normal force on the side of the web.

The normalization applied to spectral values is helpful for comparing the measured data with the calculated dispersion curves. However, this overemphasizes the high-frequency spectral components. When we replotted $H$ without normalization (**Figure 11**), it became clear that most of the spectral content was in the very low-frequency–low-wavenumber range.

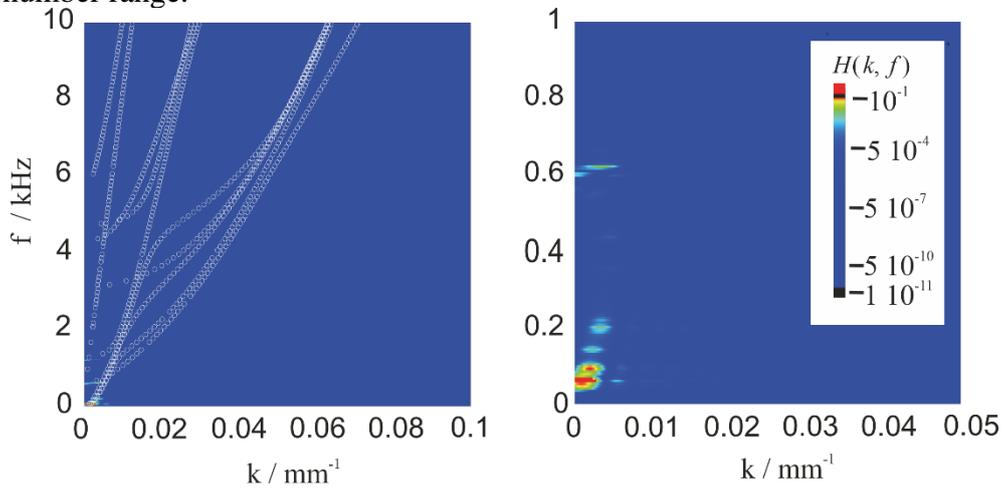

**Figure 11.** Re-plot of color-coded spectral data $H$ for ML5. Left: The same spectral range as in Figure 10; Right: Reduced spectral range. An identical logarithmic color coding has been used in both the graphs.

The dispersion curves up to 900 Hz calculated using SAFE are depicted in Figure 12. Only three modes propagated in the frequency band below 600 Hz. Based on their shapes, they could be characterized as horizontal bending (mode 1), torsional bending (mode 2), and vertical bending (mode 3). Above 600 Hz, a mode characterized by bending deformation of the web cross section (mode 4) was also be observed.

We are interested in mode 3, which corresponds to vertical bending. In this frequency range, the group velocity increased monotonically with an increase in frequency.

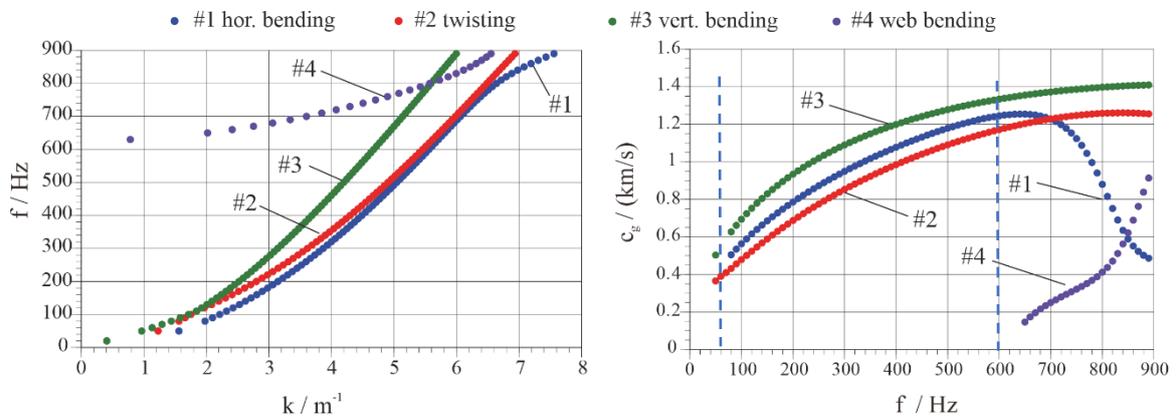

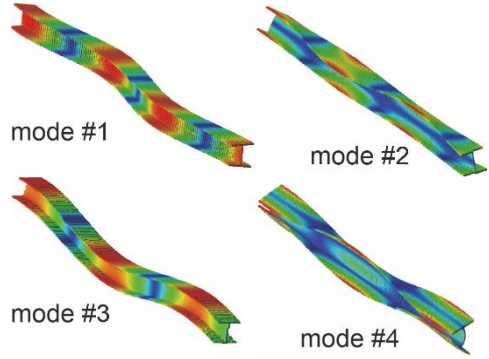

**Figure 12.** SAFE results for $f \leq 900$ Hz. Top: two display types of dispersion diagrams of the four lowest modes ($c_g$ is group velocity); Bottom: corresponding mode shapes.

### *Damage detection by displacement amplitude projection*

Figure 13 shows the measured y-displacement $|u_y(z,t)|$ over distance $z$ along ML5 and over time $t$. The line "T1" of the first arrival can obviously be assigned to the guided wave mode with the highest group velocity. There are two possible explanations for the high-intensity values appearing significantly later than the first arrival. They could either belong to wave modes with a considerably lower group velocity or be caused by multiple reflections at the ends of the sample. For the first explanation be true, they should be on a line connected to the origin of the plot, which is not the case. Therefore, we assigned them to multiple reflections and tried to eliminate them.

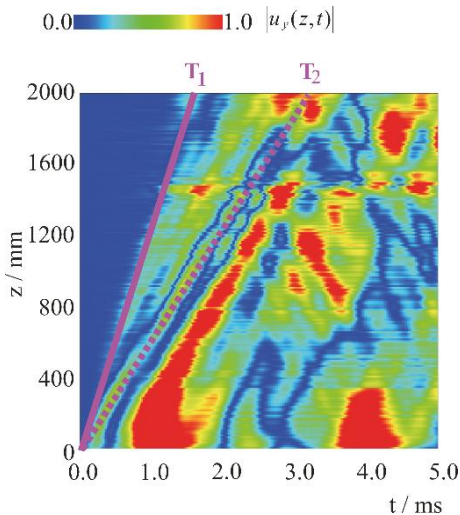

**Figure 13.** Top: Color-coded displacement $|u_y(z,t)|$ for ML5.
Below that: $|u_y(t)|$ at three selected $z$ values (indicated at the top by the dotted white lines).

The slope $(1/c_{g,h})$ of the first arrival dependence $T_1(z) = (1/c_{g,h})z$ is the inverse of the highest relevant group velocity $c_{g,h}$. Its value determined from graph $c_{g,h} = 1330$ m/s corresponds, according to Figure 12, to the group velocity of the fastest mode (#3) at $f = 600$ Hz.

The line between the area of the just forming guided waves and the area to be excluded from multiple reflections should be defined by the slowest guided wave that is relevant. Figure 2 shows that the power spectrum of the hammer excitation dropped significantly below 60 Hz. We assume that the frequency is the lower limit of the guided wave propagation. Concentrating on mode #3, we obtain $c_{g,l} = 550$ m/s. The corresponding line $T_2(z) = z/c_{g,l}$ is shown in Figure 13 as a dotted line.

The two pink solid and dashed lines define a position-dependent time window $(T_1(z), T_2(z))$ with $T_1(z) = z/c_{g,h}$ and $T_2(z) = z/c_{g,l}$, respectively, which encloses the first arrivals of guided mode #3 (vertical bending) with frequencies between $f = 60$ and 600 Hz. However, we cannot rule out the possibility that the first arrivals of other bending modes are also included in this time window.

For a quantitative evaluation, we define an intensity

$$I(z) = \max_{T_1(z) \le t \le T_2(z)} |u(z, t)| \tag{16}$$

describing a projection of the measured values $u(z, t)$ to $z$. Therefore, the proposed method is similar to the maximum intensity projection (MIP) [35] used in medical imaging and also to the C-scan imaging in ultrasonic testing [36]. In contrast to C-scans, here the width of the time window changes with the position. We call the proposed method MIP.

MIP can be easily applied to all measuring lines, ML1 to ML8. The result for our specimen is shown in Figure 14. The MIP image, with intensity normalized to its maximum, is plotted color coded over the object. A high intensity at the position of the artificial defect (thinning) is observed., where $I$ was normalized to its maximum value and displayed color-coded over the surface of the object. The intensity was significantly enhanced in the thinned part of the specimen.

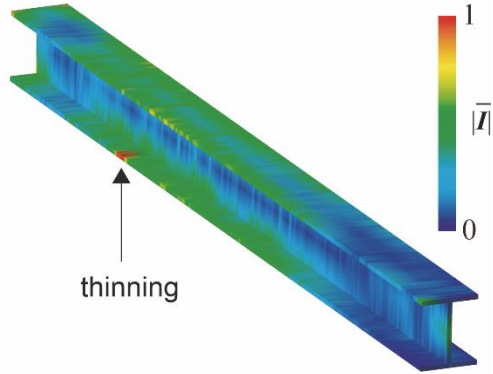

**Figure 14.** The MIP image, with intensity normalized to its maximum, is plotted color coded over the object. A high intensity at the position of the artificial defect (thinning) is observed.

The proposed MIP method for defect/damage detection can be summarized as follows:

(1) Identify the relevant defect/damage (in this study, thinning) and its position in the object of interest and study its effect on guided wave propagation. Select the modes that are sensitive to the defect/damage.

(2) Select a frequency range in which the selected mode group velocity has, at best, no overlap with other modes or only with a limited number of other modes.

(3) Define a position-dependent time window $(T_1(z), T_2(z))$ with $T_1(z) = z/c_{g,h}$ and $T_2(z) = z/c_{g,l}$ and obtain the maximum intensity over that window $I(z)$ using Eq. (16).

(4) Plot $I(z)$ for all MLs on the surface of the object to visualize the defect/damage indications.

## Summary and conclusions

Many studies have suggested reciprocity between two measurement situations: the elastic wave excitation at point A and measurement at point B and vice versa. The same signal should be received in both the situations. Often, no further conditions are mentioned but only reference to the literature [20] is made; however, the quoted statements cannot be found.

In contrast, we explicitly demonstrated that measurement reciprocity is valid under our experimental situation. This was proven by specifying the general reciprocity theorem. In this way, it became clear which sending and receiving quantities are reciprocal to each other and which other prepositions must be assumed. We presented a rather straightforward derivation step-by-step, as we believe that the interested reader could use some intermediate steps to start their own derivation for a different experimental situation. Finally, we experimentally demonstrate reciprocity and show how non-reciprocal behavior is obtained if different quantities are used.

It is nontrivial that reciprocity is still valid when the specimen is supported by point contacts. Nevertheless, we demonstrated the validity of this type of support. Proof of reciprocity for more complex support structures is desirable and might be possible. However, this was beyond the scope of this study.

Based on verified reciprocity, we demonstrated that wave motion can be easily visualized by exciting at a grid of points instead of measuring at that grid. A test sample with thinning as an artificial defect was used to demonstrate this.

The wave motion shortly after excitation shows the separation of the wave packages owing to their different propagation velocities. Multiple reflections occur at the end of the beam, forming natural-wave modes. The corresponding vibration shapes were then visualized. In the second vertical bending mode, the defect area appeared as an area with significantly increased displacement.

However, simple structures are rather an exception in the field. This implies that we cannot rely on the eigenmodes of the structure. Therefore, we propose a method, called MIP, based on the extraction of a corresponding guided mode sensitive to the defects. This guided mode was evaluated immediately after excitation. When MIP is applied to real structures with many beam branches, the early detection after excitation ensures that the guided modes reflected from the branch points do not interfere with the display. Thus, this method can be applied to branched beam structures.

Other known methods for defect detection with guided waves are based on a transformation of the space-time measured data into the wavenumber-frequency space [37], [38]. Additional guided modes are identified which appear in the defect area. For clear traces of such additional modes in the frequency-wavenumber space a sufficient number of sample points in the defect area are necessary. They are normally obtained by Scanning Laser Doppler Vibrometry. We have only three sample points inside the thinning and two on its two edges, which is why we consider the Fourier space method unsuitable in our case. However, in other cases, such as larger defects, the Fourier method could also be used in combination with data from reciprocal measurements.

In this work we describe idea of the MIP method as one possible application for data obtained by reciprocal measurements. This method is demonstrated here only for a thinning as a simple defect model. We are currently working on the demonstration of MIP also for other types of defects and for defect sizing.

## Acknowledgments

This work was supported by JSPS through a grant-in-aid for scientific research (15K01226) and the Invitational Fellowships for Research in Japan (Fellowship ID: S20084). We gratefully acknowledge this support.